\documentclass[10pt,journal,compsoc]{IEEEtran}

 \usepackage{amsmath,amssymb,amsfonts}
\usepackage{graphicx}
\usepackage{textcomp}
\usepackage{epsfig}
\usepackage{listings}
\usepackage{diagbox}

\usepackage{url}
 \usepackage{indentfirst}
\usepackage{multirow}
\usepackage{dsfont}
\usepackage{graphicx,subfigure}
\usepackage{array}
\usepackage{mathtools}
\usepackage{algorithmic}
\usepackage{balance}
\usepackage{seqsplit}
\usepackage{pifont}
\usepackage{listings}
\usepackage{tikz}
\usepackage{enumitem}

\usepackage{amsmath}
\usepackage{amssymb}
\usepackage{graphicx}
\usepackage{color}

\newcommand{\ignore}[1]{}
\newcommand{\nop}[1]{}
 \newcolumntype{P}[1]{>{\centering\arraybackslash}p{#1}}
  \usepackage[ruled]{algorithm2e}
 
\ifCLASSOPTIONcompsoc
  \usepackage[nocompress]{cite}
\else
  \usepackage{cite}
\fi
 
\ifCLASSINFOpdf
  
\else
   
\fi

\begin{document}
 
\title{Onionchain: Towards Balancing Privacy and Traceability of Blockchain-Based Applications}

\author{Yue Zhang, 
        Jian Weng, ~\IEEEmembership{Member,~IEEE}
        Jiasi Weng,
        Ming Li,
        and~Weiqi Luo,
\IEEEcompsocitemizethanks{\IEEEcompsocthanksitem
Yue Zhang, Jian Weng, Jiasi Weng, Ming Li and Weiqi Luo are with the College of Informatin Science and Technology / College of Cyber security, National Joint Engineering Research Center of Network Security Detection and Protection Technology, and Guangdong Key Laboratory of Data Security and Privacy Preserving, Jinan University, Guangzhou, 510632, China;

\IEEEcompsocthanksitem Jian Weng is the corresponding author : cryptjweng@gmail.com  ;}
\thanks{Manuscript received XX,XXXX; revised XX,XXXX}}
 
\markboth{Journal of \LaTeX\ Class Files,~Vol.~14, No.~8}%
{Shell \MakeLowercase{\textit{et al.}}: Bare Demo of IEEEtran.cls for Computer Society Journals}
 
\IEEEtitleabstractindextext{%
\begin{abstract}
With the popularity of Blockchain comes grave security related concerns.
Achieving privacy and traceability simultaneously remains an open question. Efforts have been made to address the issues, while they may subject to specific scenarios. This paper studies how to provide a more general solution for this open question. Concretely, we propose Onionchain, featuring a suite of protocols, offering both traceability and privacy. As the term implies, our Onionchain is inspired by Onion routing. We investigate the principles of Onion routing carefully and integrate its mechanism together with Blockchain technology. 
We advocate the Blockchain community to adopt Onionchain with the regards of privacy and traceability. To this end, a case-study of Onionchain, which runs in the context of Vehicular Ad Hoc Networks (VANETs), is proposed, providing the community a guideline to follow. 
Systematic security analysis and extensive experiments are also conducted to validate our secure and cost-effective Onionchain. 
\end{abstract}

}

\maketitle

\IEEEdisplaynontitleabstractindextext
 
\IEEEpeerreviewmaketitle

\section{Introduction}
 
 \IEEEPARstart{O}{ver} the past few years, Blockchain has drawn significant attention from both academy and industry \cite{pearce2019blockchain}. Blockchain is a novel paradigm where distrustful parties make transactions and manage data without involving a trustworthy third-party. Here transactions refer to interactions occurred between these parties. Blockchain achieves tamper-resistance and traceability for the transactions, offering anonymity and decentralization for the parties. 
 Due to these advanced features, Blockchain can
be applied into a wide spectrum of applications, ranging from cryptocurrency, financial services, crowd-sourcing systems \cite{li2018crowdbc,liu2014toward}, and Vehicular Ad Hoc Networks (VANETs) \cite{sharma2017block,tzeng2015enhancing}. 
According to a report from Meticulous Research, the global Blockchain market will hit \$28 billion by 2025 \cite{blockchianmarket}.\looseness=-1

 However, along with its popularity, Blockchain has come an increasing number of attacks, severely undermining the victim's security and privacy. In regards to privacy, although Blockchain can provide anonymity innately, it subjects to various cyber-attacks. For example, Fergal \textit{et al.} \cite{reid2013analysis} show that an attacker may disclosure the real identity of a given victim by analyzing his public transaction history. Efforts have been made to counter these attacks. An example that addresses the privacy concern is HAWK proposed by Kosba \textit{et al.},  storing the encrypted transactions instead of plain-text ones so that the transactional privacy is guaranteed \cite{kosba2016hawk}. The term privacy in our paper refers to data privacy unless explicitly stated otherwise. In data privacy schemes, the identities of parties are public. However, given a message, no efficient adversary can determine if the message is from a specific party.

 While achieving privacy, these solutions may fail to provide traceability, hindering them to adapt to some scenarios. 
 For example, in crowd-sourcing systems (or crowd-sensing system in VANETs \cite{wen2014quality}), employers may release tasks for employees, while employees choose the tasks of interest and get paid when they offer proper solutions. During the life-cycle of a task, Blockchain endorses the behaviors of both employers and employees, so that employers/employees can be held accountable when malicious behaviors occur. Offering the privacy blindly in such a scenario will lead a malicious employers/employees to evade responsibilities.

 The fact that Blockchain fails to provide traceability and privacy simultaneously impedes the progress of its deployment. To address the issue, there are a few existing works. For example, the primitive, linkable group/ring signatures \cite{chaum1991group,bresson2002threshold}, are possible solutions for this issue. In these manners, a signature can be generated anonymously to present the willing of the parties in this group, and the group manager can reveal the identity/identities that perform the signature generation. However, the drawback is also obvious, since the group manager can be malicious, leading the abuse of the identity disclosure. Lu \textit{et al.} propose ZebraLancer \cite{lu2018zebralancer}, where a new primitive common-prefix-linkable is presented to realize a trade-off between traceability and anonymity. In the context of crowdsourcing, a malicious employee can submit his effort twice, to gain a doubled reward. ZebraLancer is capable of identifying the double-submission problem and tracing the dishonest party. However, ZebraLancer is not omnipotent, since it fails to discuss how to deploy their protocol in other scenarios where there is no such malicious behavior like double-submission. Liu \textit{et al.} \cite{liu2019anonymous} achieve the traceability and privacy simultaneously by introducing an identity management entity (IDM) and let IDM recover the identity of the misbehaving party. While the problem solved, their design partly violates the discretization of Blockchain. Other solutions, such as \cite{miller2017empirical}, may also subject to specific scenarios. Therefore, balancing the traceability and privacy simultaneously remains an open question.

 This paper tries to provide a general solution for this issue. To this end, we propose Onionchain, offering traceability and privacy at the same time. 
 The intuition behind our research is that we observed that Onion routing can achieve traceability and privacy in different contexts. Onion routing \cite{goldschlag1999onion} is an infrastructure designed for anonymous communication. It uses a set of onion routers instead of using regular routers. The onion routers encrypt and relay packets between a source node and a destination node. 
 In  terms of privacy, onion routers resistant to both eavesdropping and traffic analysis innately, since the encryption process is present. In  terms of traceability, Onion routing is a communication mechanism featuring routing and packets transmitting. In the view of the onion routers, they can identify an intended source and an intended destination by decrypting the encrypted packets accordingly.  But, for a single onion router, it has limited routing information and provides packets forwarding partly. Therefore, for most of the time, as a destination or an onion router, they cannot trace the packets back to the source. This is because tracing a packet needs the efforts of all relaying onion routers, but the onion routers will not work cooperatively to make this happen since they are different nodes in the network.  \cite{goldschlag1999onion}
 Inspired by these two features, we systematically investigate the underlying principles of the Onion routing and design Onionchain. 
 
 The high-level idea is that we introduce Blockchain as a trustworthy party, and enable the onion router-like nodes to decrypt the packets to Blockchain conditionally. That is, Blockchain will perform the identity disclosure according to the willing of majority. In such a way,  Onionchain achieves traceability and privacy simultaneously: Normally, Onionchain will offer privacy for each party. In the special case where disclosure is needed, parties will work closely to make the disclosure happen.
 To notice, our Onionchain integrate the mechanism of Onion routing together with Blockchain technology instead of using Onion routing as a building block directly. Using Onion routing with no change is fairly trivial. More importantly, while achieving privacy, Using onion routing barely fails to provide a proper solution for nodes to work jointly to disclose a specific identity \cite{reed1998anonymous}.    \looseness=-1
 
We advocate the community to adopt Onionchain when Blockchain applications require achieving both of traceability and privacy. As a case study, we deploy our Onionchain in the case of VANETs. In such a context, vehicles by sharing the information of the road environment work jointly to prevent traffic jams and accidents. Therefore, it is crucial for vehicles to obtain information without any errors in a timely manner. 
Blockchain provides a decentralized data sharing environment for VANETs, as shown in prior efforts \cite{singh2017blockchain,kang2018blockchain,knirsch2018privacy}, offering privacy-preserving for vehicles. However, a malicious vehicle may poke the pitfalls of VANETs in various ways, spreading false information intentionally, misleading other vehicles. One example is Blockhole attack \cite{kurosawa2007detecting}, where a malicious vehicle pretends to be another benign vehicle, and discard any packets that flow toward it, causing severe packet loss. 
In such a context, providing privacy without traceability, such as what the traditional Blockchain-based solutions have done, is not enough.
As a countermeasure, we introduce our Onionchain to guide behaviors of vehicles. Our Onionchain can provide the privacy-preserving for a benign vehicle, and identify the malicious vehicle when an attack occurs.  \looseness=-1

Major contributions can be summarized as follows:
\begin{enumerate}
\item We design Onionchain, featuring a suite of customized protocols and algorithms, achieving traceability and privacy simultaneously. 
\item We demonstrate our Onionchain in the context of vehicles communication systems, avoiding vehicles to spread false information or discard valuable information intentionally. For the sake of generality, the demonstration in our paper can be trivially extended to other similar scenarios. 
\item We validate our Onionchain and the proposed countermeasure for VANETs. We prototype the countermeasure on Ethereum. Excessive experiments are performed to evaluate our implementation, showing the overload is fairly acceptable.

\end{enumerate}

The rest of our paper is organized as follows. In section~\ref{sec2:background}, we introduce primitives involved in our paper, including Blockchain, smart contracts, Onion routing, and VANETs.  In Section~\ref{sec:threatmodel}, we present the security model of Onionchain. In Section~\ref{sec:torchain}, we elaborate on the design criteria of Onionchain and deploy it in the context of the vehicles communication system. We systematically analyze the security of Onionchain in Section~ \ref{sec:secanalysis} and evaluate Onionchain experimentally in Section~\ref{sec:eva}. Related works are reviewed in Section~\ref{sec:relatedwork}. We conclude the paper in Section~\ref{sec:con}.

\section{Background}
\label{sec2:background}

In this section, we first provide a brief introduction of Blockchain and smart contracts technology. Afterward, we elaborate an overview of onion routing to demonstrate the principles behind. We also show the architecture of Vehicular Ad Hoc Networks (VANETs).

\subsection{Blockchain}

Blockchain was first proposed by Nakamoto in \cite{nakamoto2008bitcoin}, as the fundamental primitive of decentralized digital currency, Bitcoin.  In the context of Bitcoin, Blockchain is a distributed public ledger, which is initially designed for mutually distrustful parties to make transactions without involving a trustworthy center.  
To achieve this, each party needs to maintain the entire copy of Blockchain and works cooperatively to record the transactions. 
The transactions are organized as a data block, which is composed of a block header and a block body. Basically, the transactions are stored in the block body, while the block header contains a root hash and a reference.
The root hash is the hash of a Merkle Tree \cite{szydlo2004merkle}  of all transactions in the block body. The reference, which is also a hash value, is computed from the block header of the previous block. 

Each party has an identity, termed address, which uniquely refers to a specific party. To obtain an identity, each party needs to generate a pair of public/private key. The private key is used for signature verification, and the public key is used for signature generation and builds the address.  
To notice, not all parties can record transactions on Blockchain. The one who has the priority to append data, termed miner, needs to provide a solution to a certain puzzle, known as proof-of-work ((PoW) \cite{gervais2016security} or proof-of-stake (PoS) \cite{king2012ppcoin} etc. For example, the PoW is a hash computing puzzle that can only be solved when an appropriate answer is fed into the hash function. Due to the property of the hash function, the puzzle requires much efforts to find out such an answer, while the validity of the answer can be trivially checked.

 Blockchain records all transactions by default, and it achieves tamper-resistance and traceability innately. The main reasons are: (i) For a single block, each block maintains the hash of a Merkle Tree that computed from transactions in the block body, so that no more modifications can be made once the block is generated, since changes will make the hash root of Merkle Tree fail to match the original one; (ii) For the entire Blockchain,  Blockchain is maintained by each party, so that it is impossible for a single party to modify it, since modifying the Blockchain of himself will not make any influences to other parties. Moreover, due to the reference to the previous block, the integrity of the entire Blockchain is also guaranteed, since each party can verify it chronologically without any changes; (iii) The only way to make a modification happen, i.e., adding a new transaction to Blockchain, may require a lot of efforts since the one who wants to do so must solve the puzzle first. 
 
 \subsection{Smart Contracts}
 \label{subsec:sc}

Smart contracts are digital agreements that are made between different parties. The term was first introduced in 1994 by Nick Szabo \cite{szabo1997formalizing}, and gained many attentions with the booming of Blockchain. 
In smart contracts, instead of using the printed contracts, and enforcing to execute via a suite of regulations or laws, smart contracts are grouped into lines of codes, and will be executed when specific conditions are fulfilled. To ensure the proper execution of the codes, a secure environment must be guaranteed. Otherwise, an attacker may compromise the codes to maximize his benefits. This was a giant barrier that hinders the development of smart contracts. However, as mentioned earlier, Blockchain contributes a lot to the popularity of smart contracts, since decentralized Blockchain networks and other advanced features of Blockchain can offer such a secure environment for smart contracts. 
Examples of Blockchain that run smart contracts  include Ethereum \cite{wood2014ethereum} and Hyperledger \cite{cachin2016architecture}.

 \subsection{Onion Routing}
 \label{subsec:onionrouting}

 Onion Routing \cite{goldschlag1999onion} 
  is an infrastructure for anonymous communication. Instead of creating socket connections between the two communicating machines directly,  onion routing achieves the communication via a sequence of machines termed onion routers. Onion routers relay the two communicating machines, offering routing and addressing, allowing the two communicating machines to remain anonymous in one way or both ways.  One application of onion routing is anonymous web browsing. In this case, a user may want to browse a website without disclosing his identity to the webserver. That is, anything related to him, such as IP address or MAC address, shall be removed from the browsing request.  

\begin{figure} 
\centering
\includegraphics[width=0.5\textwidth]{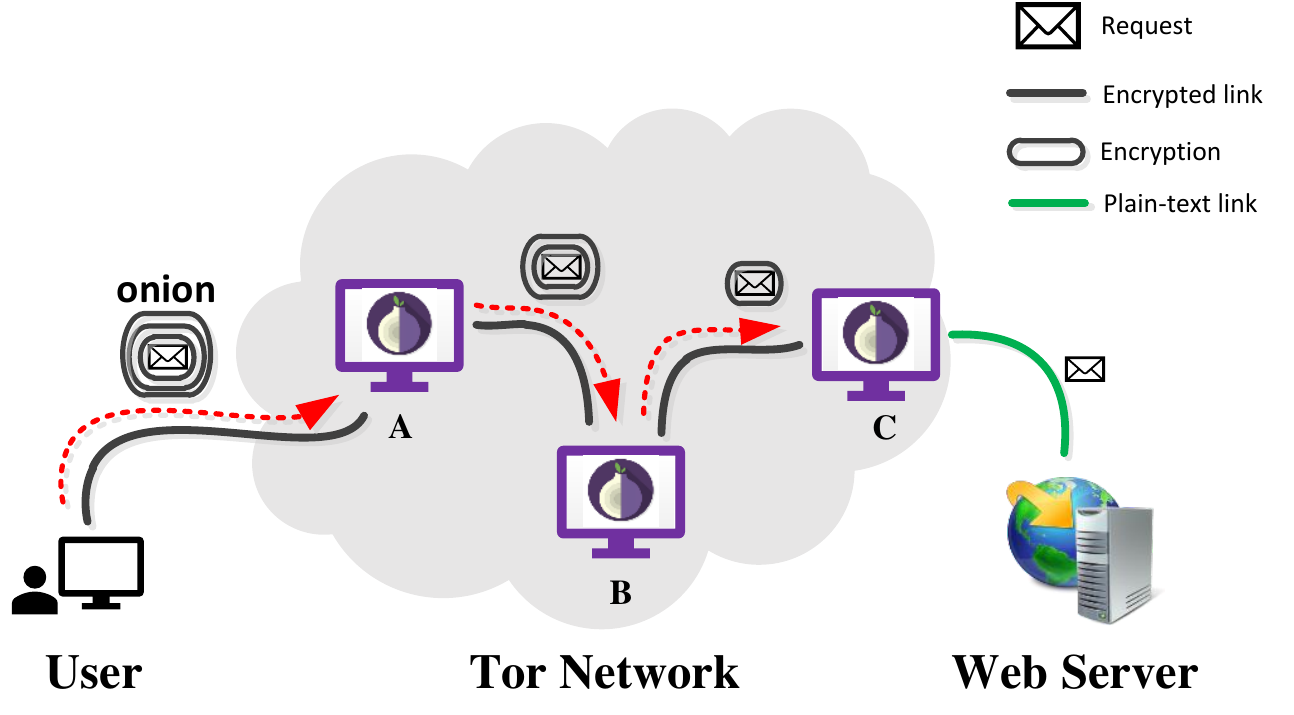}
\caption{\label{fig:tor}  The principle of onion routing }
\end{figure}

To demonstrate the principle of onion routing, Fig.~\ref{fig:tor} illustrates the process of anonymous web browsing. Specifically, when the user wants to browse a website anonymously, he must follow the steps elaborated below: 
\begin{enumerate}
\item Initially, the user uses an onion proxy to randomly choose three onion routers ($A, B, C$ in the figure) in the network and negotiate three different keys with the three onion routers.   We denote the encryption keys as $Key_{User-A}$, $Key_{A-B}$, $Key_{B-C}$, where the first letter in subscript refers to the sender of the packet, while the second letter refers to the responder of the packet.  To notice, we are aware that the public cryptography and other technologies/roles are also involved in this process, and we will not discuss the key negotiation in detail for the sake of brevity. 
\item To achieve anonymous communication, the user encrypts the browsing request with the encryption keys successively. The user also hard-code the next-hop information inside the encrypted browsing request, to make sure that each onion router is aware of its next hop.  The encrypted browsing request is called the onion.
\item The user sends the onion to the first onion router $A$. 
The first onion router removes the first layer encryption by decrypting the onion with the key $Key_{User-A}$, which is negotiated previously. Once decrypted, the onion router then knows which onion router is the one it should send the onion to. In our context, the first onion router will send the onion to $B$.
\item $B$ receives the onion and removes the second layer encryption by decrypting the onion with the key $Key_{A-B}$, so that it obtains its next-hop information. The second onion router then sends the onion to $C$. 
\item $C$ removes the last layer of the onion by decrypting the onion with the key $Key_{B-C}$, and sends the browsing request, which is in plain-text, to the intended web server. The three onion routers then communicate with each other and work jointly to relay the packets between the user and the webserver.
\end{enumerate}
 
 Since the links between each other are encrypted, each router may only obtain limited information about the connection. For example, for the onion router $A$, it only knows who is the sender of the packets and who is its next hop.  For the onion router $C$, it only knows the information about the webserver and its previous hop. For the onion router $B$, it neither knows the information about the user nor the information about the webserver. The only information he obtained is the hop information.  In such a way, onion routing achieves anonymous communication. 

 
\subsection{Vehicular Ad Hoc Networks (VANETs)}

Vehicular Ad Hoc Networks (VANETs) \cite{sheikholeslam1993longitudinal,weng2018benbi,zander1992distributed}, as the term itself implies, is designed for vehicles to achieve communication. It has aroused researchers’ interest since it offers a paradigm for facilitating smart city. In vehicles communication systems, vehicles are equipped with a set of sensors,   Global Positioning System (GPS) \cite{willson2008gps}, and Radio Frequency Identification (RFID) tags \cite{finkenzeller2010rfid}, which allows them to sense the environment around. Examples of the applications of vehicles communication systems include navigation, path planning. For instance, in the case of path planning, vehicles may communicate with each other to share the traffic information, so that a suitable route can be selected, avoiding the traffic jams\cite{robertson1991optimizing}.  

\begin{figure} 
\centering
\includegraphics[width=0.5\textwidth]{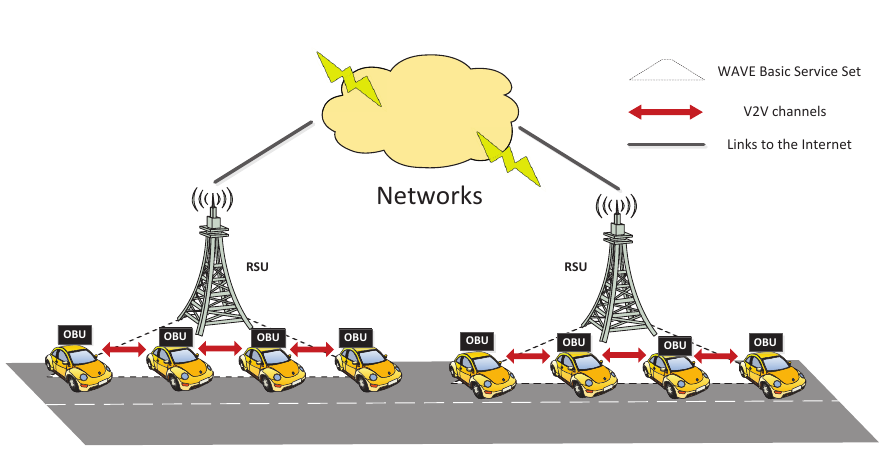}
\caption{\label{fig:v2v} Overview of vehicles communication system }
\end{figure}

Fig. \ref{fig:v2v} shows an overview of the vehicles communication system. Instead of using traditional Wireless Local Area Network (WLAN) and the wired network, vehicles communication system introduce Road-Side Units (RSUs) to connect the internet. These RSUs are installed along the road-sides, providing Internet access for vehicles. The coverage of each RSU is termed Wireless Advanced Vehicle Electrification (WAVE) Basic Service Set. Basically,  vehicles connect to the RSUs via their  On-Board Units (OBU), which is a component mounted inside each vehicle.  When there is no RSU available,  the vehicles can also achieve local area communication.

\section{Security Model}
 \label{sec:threatmodel}
In this section, we first present the security goals of Onionchain. We will also make the basic assumptions for our design afterwards.

\subsection{Security Goals}

We claim the ultimate goals of Onionchain. It can be observed from the introduction and the principle of Blockchain that the current design of Blockchain applications fails to achieve privacy and traceability simultaneously. Therefore, Onionchain has the following essential design requirements: 
  
\textbf{Privacy}: First of all, the privacy issue is our main concern. The term privacy refers to data privacy, where identities of parties are public, while there is no efficient adversary can determine if a given message is sent from a specific party. That is, as long as a party behave honestly, no body can link the message he sent to his public identity. The identity may include IP address, Mac address, or other personal information. Onionchain will provide a countermeasure for these parties, against their privacy being leaked.

\textbf{Traceability}: Second, with regard to traceability, Onionchain will disclose a specific identity of a party following the willing of the majority. That is, when a dishonest/malicious behavior occurs, the majority of parties may require the Onionchain to reveal the identity of a party, and Onionchain will fulfill the requirement. 
 
\subsection{Assumptions}

In our system, an attacker's goal is to reveal the identities of parties and pokes the design flaws of our Onionchain.
Without loss of generality, we make the following assumptions for attackers and our system. 
\begin{enumerate}
\item In our design, cryptographic mechanisms will be involved, and we assume these cryptographic mechanisms are not a source of vulnerability. For example, an attacker cannot break the AES encryption or forge a digital signature. 

 \item We assume that the fundamental assumption of Onion routers still holds. That is, the nodes in the network are randomly distrusted. Therefore, when a party chooses nodes from network randomly, the chosen nodes are not all manipulated by an attacker.  

\item We assume that an attacker can access all the transaction history without any change. This is reasonable since Blockchain is a public ledger that is free to download for anyone. This assumption offers the worst scenario for our design. Our design can hinder the privacy leakage in this scenario, and it also hinders other scenarios.

\item We assume that a large number of parties are involved in our system. This is reasonable since systems, such as vehicles communication systems, may involve hundreds of participants.

\item A malicious party may spread false message intentionally, misleading other parties in various ways. We assume that after after a message being spread, other parties have ways to identify whether the information is a false one. For example, false information can be fake traffic information on a specific road, which can be trivially identified when a vehicle arrivals there.  However, more ways of identifying false information are out of our focus, and we will not discuss it. 
 
\end{enumerate}

\section{Onionchain}
\label{sec:torchain}

We first explain our insights behind our design. 
From the introduction of onion routing in Section \ref{subsec:onionrouting}, we learn that onion routing resists to both eavesdropping and traffic analysis innately. We also observe that the onion routers will relay the packets between an intended source and an intended destination. For most of the time, as a destination, it cannot trace the packets back to the source. This is because that the destination and its previous hop, which is an onion router, are different nodes in the network. They shall not share the hop or routing information with each other. But what if the destination node and the three onion routers are all manipulated by a third-party behind? In this case, the third-party can obtain all the routing information, and trace from a destination back to a source. That is, he is capable of disclosing the identity of the source. 
Here, such a third party can be a good or evil, depending on if and how he will perform the disclosure.
If he is ultimately fair and only behaves according to the willing of majority, he is a good one. However, does such a flawless third party exist? Actually, Blockchain can be such a third party. Motivated by this, we have designed a suite of protocols, termed Onionchain which make Blockchain to be such a third party. In the following of this section, we will present the criteria of our design in detail. 

\nop{
\begin{figure} 
\centering
\includegraphics[width=0.5\textwidth]{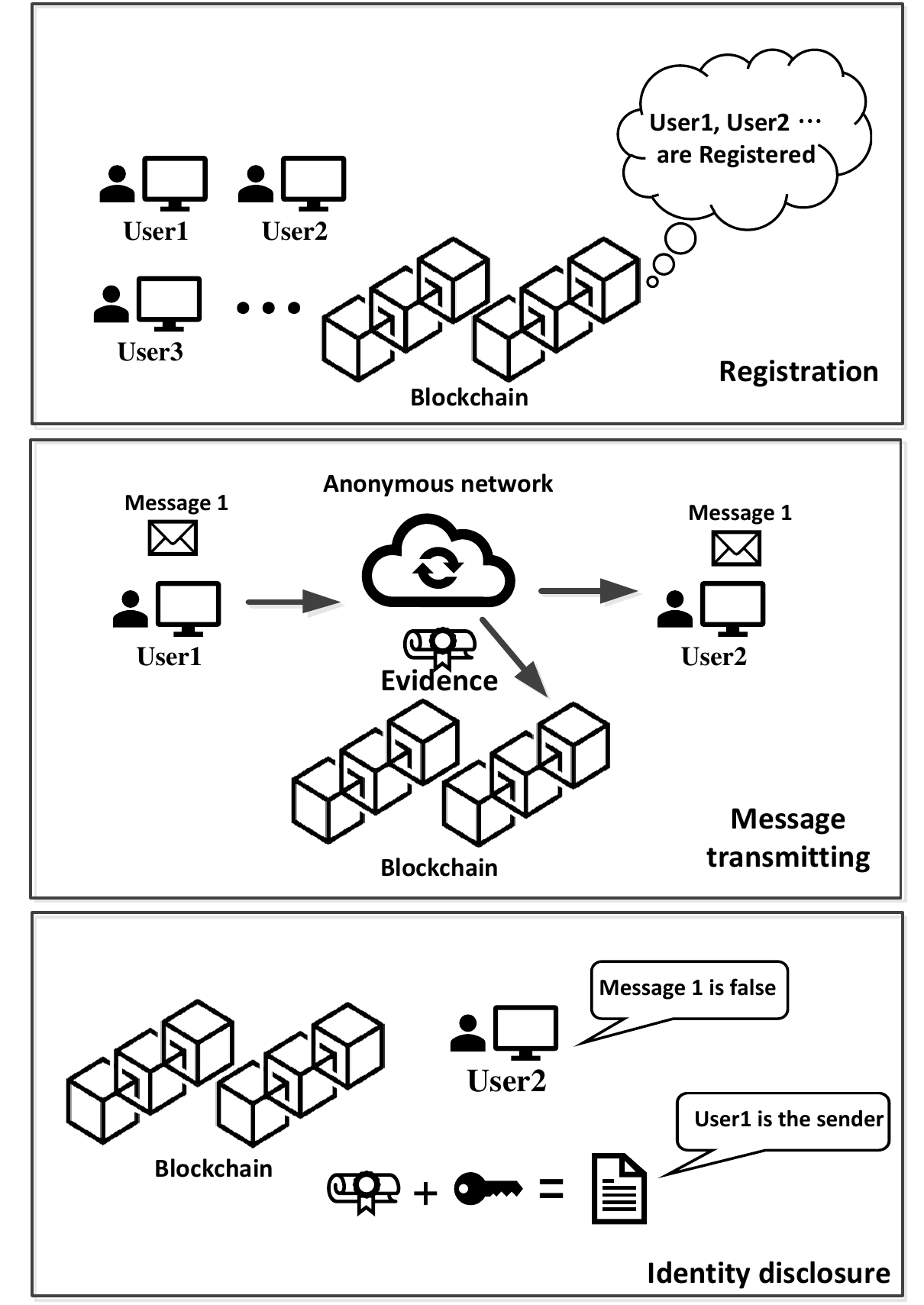}
\caption{\label{fig:overview} Overview of Onionchain }
\end{figure}
}
\subsection{Overview}

Onionchain offers three core protocols for parties to achieve privacy-preserving and tracing transactions, including the registration,  message transmitting, and identity disclosure. Registration is used for parties to join Onionchain. To this end, each party must provide their real identity to Onionchain distributedly, and Onionchain will store these information onto Blockchain. All these information are free to access for the public. In regard of privacy-preserving, although the information is publicly accessible, an attacker cannot link a specific message to its sender according to our design. The privacy of each party is still preserved. Message transmitting is a protocol that defines how two parties transmit a message via the network. Different from the traditional network communication, the message sender and the message responder may need to write data onto Blockchain. The data, termed evidence in our case, is encrypted by negotiated keys, and also plays an important role in the identity disclosure protocol. The disclosure protocol will perform when the parties have the requirement to disclose a specific sender. Say, when false information is identified, the parties want to know who is the sender of this false information. In this case, our identity disclosure protocol can link the false information to a specific sender by decrypting the evidence. 


\subsection{Design Criteria}

In this section, we will elaborate on the design criteria of registration protocol, message transmitting protocol, and identity disclosure protocol. 
For ease of description, we summarize the basic operations in Table \ref{tab:notion}.

\begin{table} 
\centering 
\caption{Summary of notations}
\label{tab:notion}
\begin{tabular}{|c|P{4cm}| }
 \hline
{\bf Notation} &  {\bf Description  }     \\
\hline
  $a\rightarrow b$  &  a routing message, where $a$ is the source and $b$ is the destination \\
 \hline
  $PubK_{i}$  &  the public key of a party $i$   \\
 \hline
   $SK_{i}$  &  the private key of a party $i$   \\
 \hline
   $ s_1 || s_2 $  &  a combination of a string $s_1$ and a string $s_2$   \\
 \hline
\textbf{ENC}($key$,$msg$)  &  The AES encryption process, where $key$ refers to a encryption key, and $msg$ refers to a message  \\
\hline
 \textbf{DEC}($key$,$ctext$)  &  The decryption process, where $key$ refers to a encryption key, and $ctext$ refers to a cipher-text  \\
\hline
 \textbf{SIGN}($SK$,$msg$)  &  The signature generation process, where $SK$ refers to a private key, and $msg$ refers to a message  \\
\hline
 
 \textbf{VERIFY}($PubK$,$sig$,$m$)  &  The signature generation process, where $PubK$ refers to a private key, and $sig$ refers to a signature, while $m$ is the message  \\
 \hline
\end{tabular}
 
\end{table}

\textbf{Registration protocol}: This protocol is used for parties to sign up in our system. It works as follows:  (i) When a party $A$ wants to join the system, he first creates a public/private key pair, denoted as $SK_{A}$, $PubK_{A}$,  which can be used to perform signature generation and verification. (ii) He uses his private key $SK_{A}$ to sign his real identity $ID$, and put his public key $PubK_{A}$ and the generated signature $S$ together to generate a registration request, i.e $regReq = (PubK_{A} || S)$.
The $ID$ here uniquely refers to a specific party. 
He submits the registration request to the Blockchain P2P network and preserves his private key carefully. To notice, before the request having been approved and  written onto Blockchain, $A$ is not permitted to send any other request.  (iii)
Other parties will verify the registration request before they write it onto Blockchain. The verification process is to make sure the signature is generated from the attached public key $PubK_{A}$, and the one who sends the request has the same identity as attached. Otherwise, other parties will reject the request. The other parties also need to check if the public key attached has been used by other parties else. Since the request can be viewed by all the parties, it is trivial for them to identify a duplicate public key. That is, each party compares the attached public key with his own public key and confirms that they are not the same. If they are the same, the party with the same public key will broadcast a confliction. The confliction is also a type of request that has a higher priority for other parties to process.
Also, other parties may need to check if the identity used is a valid one or fabricated by the party itself. Therefore, the Blockchain used in our paper is a permissioned blockchain. The permissioned blockchain is a type of Blockchain that requires permission to join, and limits the parties who can provide the solution for the puzzle, i.e., being the miners. Recall that all parties other than miners can submit their transactions into Blockchain network, but only miners have the permission to record the transactions. This is reasonable because it reduces the risk of being attacked by some attacks, e.g., 51\% attack and selfish mining attack. The process of how to group the requests as transitions, and how to write the transactions are as same as Bitcoin, and we will not go to details due to the page limitation. \looseness=-1

\begin{figure} 
\centering
\includegraphics[width=0.5\textwidth]{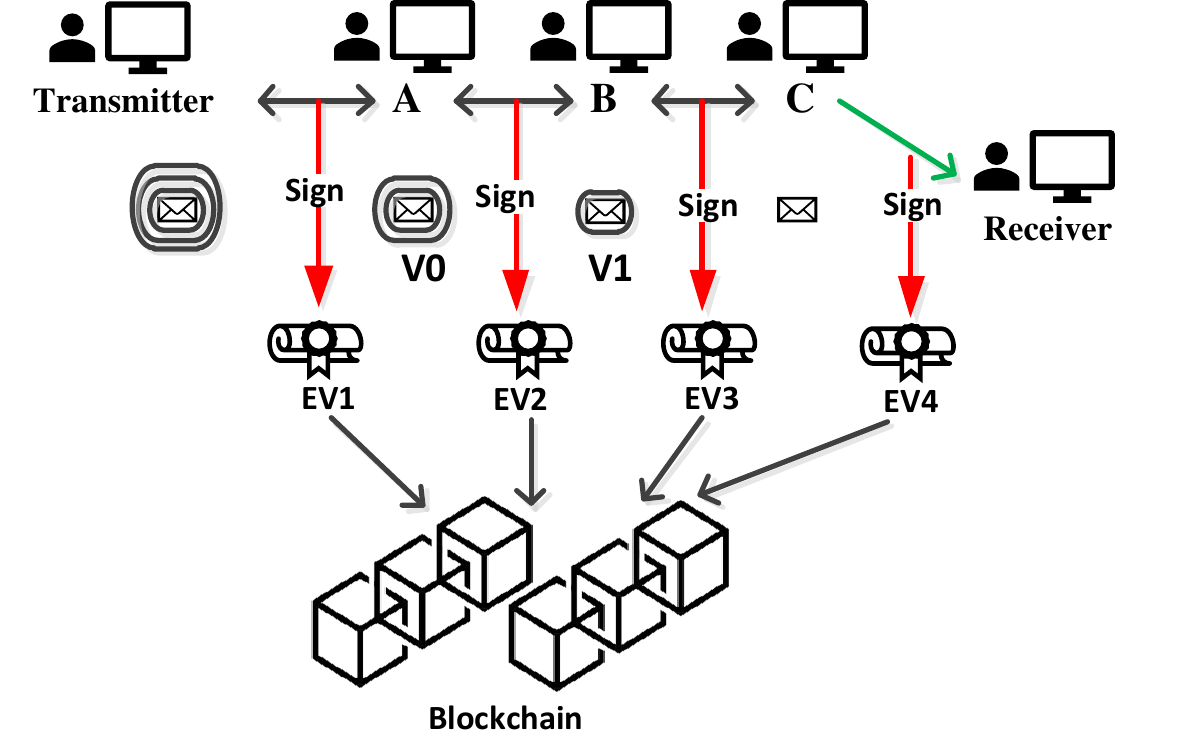}
\caption{\label{fig:message} Overview of message transmitting protocol }
\end{figure}

\textbf{Message transmitting}: This protocol is the core engine of our design. As shown in Fig.~\ref{fig:message}, we use three hops as relays between a transmitter $T$ and a receiver $R$ to demonstrate our principle. Using three hops as relays are the minimum requirement of our system. However, parties can choose more than three hops to achieve better privacy. Specifically, we elaborate on the steps below: 
\begin{enumerate}
\item Initially, the sender who wants to send the message first randomly chooses three nodes in the Blockchain P2P network. The three nodes are denoted as $A,B,C$ respectively. The sender also negotiates the three different keys with the three nodes. The key negotiation process is as same as that in onion routing, so we will not repeat it due to the page limitation. We denote the encryption keys as $K_{T-A}$, $K_{A-B}$, $K_{B-C}$, where the first letter in subscript refers to the sender of the packet, while the second letter refers to the packets responder. For example, $K_{T-A}$ is used to encrypt a message that sent from $T$ to $A$. \looseness=-1 

\item $T$ encrypts the message with the three encryption key successively. The user also hard-code the next-hop information, e.g., $A\rightarrow B$ , inside the encrypted message, to make sure that each node is aware of its next-hop respectively. As mentioned earlier, the encrypted message is called evidence in our paper. To notice, during the message transmitting, there is more than one evidence are generated. We denote this evidence $EV_{0}$. We also denote the message as $m$, which contains a timestamp to grantee the freshness of messages. Formally, $EV_{0}$ can be represented as equation (\ref{eq:ev0}) : 

\begin{equation}
\begin{aligned}
 EV_{0} = \textbf{ENC}(K_{T-A},
 (A\rightarrow B~||\textbf{ENC}(K_{A-B},\\
(B\rightarrow C~||\textbf{ENC}(K_{B-C},(C\rightarrow R~|| m))))))
\end{aligned}
\label{eq:ev0}
\end{equation}
 
\item $A$ first removes the first layer encryption by decrypting the $EV_{0}$ with the key $K_{T-A}$, which is negotiated early. We denote the decrypted onion-like packets as $V_{0}$. Formally,$V_{0}$ can be represented as equation (\ref{eq:v0}) : 
\begin{equation}
\begin{aligned}
V_{0} =  \textbf{ENC}(K_{A-B},
(B\rightarrow C~||~\textbf{ENC}(K_{B-C},~\\(C\rightarrow R~|| m))))
\end{aligned}
\label{eq:v0}
\end{equation}
Once decrypted, $A$ then knows which node is the next hop. In our context, $A$ will send $V_{0}$ packet to $B$, and the key used between $A$ and $B$ is $K_{A-B}$. Before this process, $A$ also needs to generate a new evidence and submit it to Blockchain first. Here, $A$ and $T$ are required to work closely, and generate a the new evidence $EV_{1}$. To this end, $T$ signs $EV_{0}$ with his private key, and sends $ EV_{0} | SIGN(SK_{T},EV_{0})$ to the first node $A$. The first node verifies the signature using the public key of $T$. Recall the public key was written onto Blockchain in the registration process and free to index. This step is used to ensure the message is sent from $T$, not other parties.  Once the verification processed, $A$ signs $  SIGN(SK_{T},EV_{0})$ with his own private key $SK_{A}$. Thereafter, the two parties negotiate a new key $PK_{T-A}$, termed proof key,  then encrypts $SIGN (SK_{A}, SIGN(SK_{T},EV_{0}))$ with $PK_{T-A}$ to generate the new evidence $EV_{1}$. This step ensures that $A$ receives the message successfully. The $EV_{1}$ are suppose to be written on Blockchain. Here, $T$ and $A$ have the same proof key $PK_{T-A}$, so that they can check the signatures and confirms that they all follow the protocol properly, while other parities will have no information about $EV_{1}$. To notice, the $K_{T-A}$ and $PK_{T-A}$ can not be the same, and each party is required to keep the $PK$ carefully for the further usage. Also, every time when a new message is sent, each party needs to negotiate a new proof key.  Formally, $EV_{1}$ can be represented as equation (\ref{eq:ev1}): 

\begin{equation}
\begin{aligned}
 EV_{1} ~= ~\textbf{ENC}(~PK_{T-A},\textbf{SIGN}(~SK_{A},\\ (\textbf{SIGN}(~SK_{T},~EV_{0})))))
\end{aligned}
\label{eq:ev1}
\end{equation}

\item $A$ will wait until $ EV_{1}$ is written onto Blockchain. Thereafter, $A$ will send $V_{0}$ packet to $B$. $B$ then removes the  second layer encryption with the key $K_{A-B}$ and generate the  onion-like packets as $V_{1}$.  Formally, $V_{1}$ can be represented as equation (\ref{eq:v1}): 
\begin{equation}
\begin{aligned}
V_{1} = \textbf{ENC}(K_{B-C},(C\rightarrow R~|| m)) 
\end{aligned}
\label{eq:v1}
\end{equation}
Afterwards, $A$ and $B$ work closely to generate a new evidence
$EV_{2}$. The process is similar to what we have elaborated in previous step, and we will not repeat it. Specifically,  $EV_{2}$ can be represented as equation (\ref{eq:ev2}): 

\begin{equation}
\begin{aligned}
 EV_{2} = \textbf{ENC}(PK_{A-B}, ~\textbf{SIGN} (~SK_{B},  \\  \textbf{SIGN}(SK_{A}, V_{0} ||EV_{1}))) 
\end{aligned}
\label{eq:ev2}
\end{equation}

\item $B$ will wait until $ EV_{2}$ is written onto Blockchain. Thereafter, $B$ will send $V_{1}$  to $C$. $C$ then removes the  finally layer encryption with the key $K_{B-C}$ and obtains the message $m$ in plain-text. At this time,  $B$ and $C$ work closely to generate a new evidence
$EV_{3}$.  Specifically,  $EV_{3}$ can be represented as shown in equation (\ref{equ:e3}): 
\begin{equation}
\begin{aligned}
\label{equ:e3}
 EV_{3} ~= \textbf{ENC}(~PK_{B-C},~\textbf{SIGN}(SK_{C}, \\  ~\textbf{SIGN}(SK_{B},V_{1} || EV_{2}))) 
\end{aligned}
\end{equation}

\item $C$ will wait until $ EV_{3}$ is written onto Blockchain. Afterwards, $C$ will send message to $R$. $R$ and  $C$ then work closely to generate a new evidence
$EV_{4}$. Specifically,  $EV_{4}$ can be represented as equaltion (\ref{eq:ev4}): 
\begin{equation}
\begin{aligned}
 EV_{4} ~= \textbf{ENC}(~PK_{C-R},~\textbf{SIGN}(SK_{R}, \\  ~\textbf{SIGN}(SK_{C}, m || EV_{3}  ))) 
\end{aligned}
\label{eq:ev4}
\end{equation}
  
\end{enumerate}
 
\begin{figure} 
\centering
\includegraphics[width=0.5\textwidth]{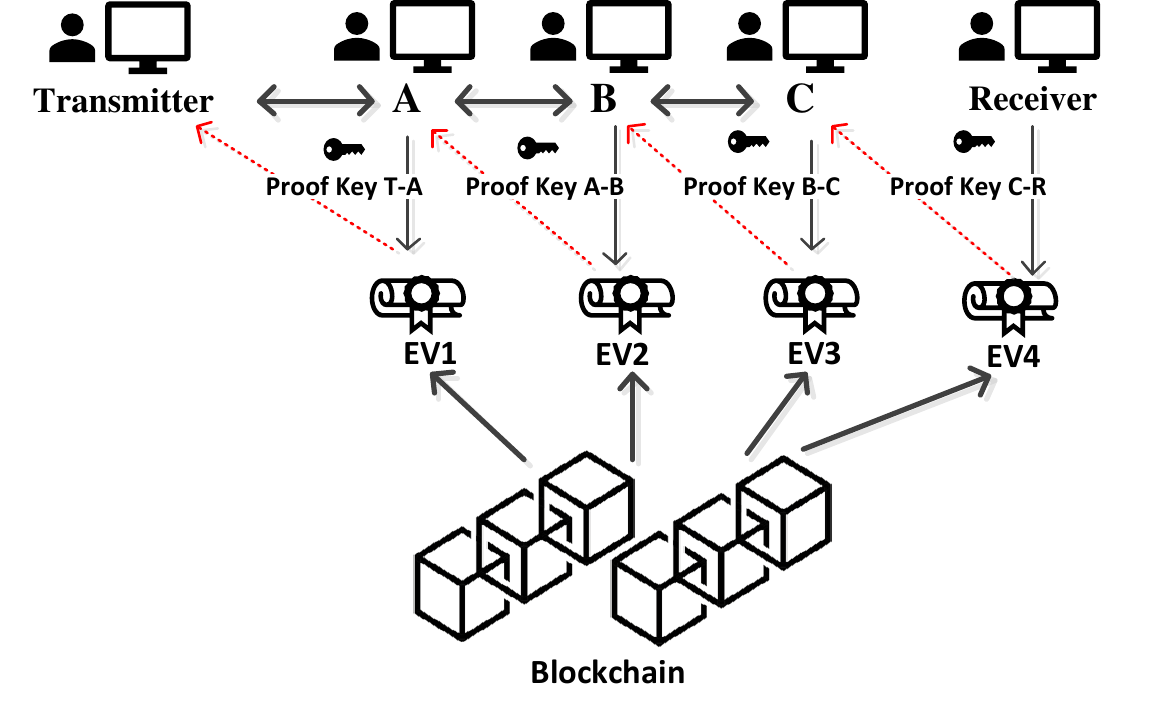}
\caption{\label{fig:trace} Overview of identity disclosure protocol }

\end{figure}

\textbf{Identity disclosure}: 

Identity disclosure is available when a dishonest behavior has been detected.
 As shown in algorithm.~\ref{alg:2}, we elaborate the entire process of Identity disclosure. 
Basically, it is the reverse process of message transmitting. To perform an identity disclosure, the parties that involved in the message transmitting may need to decrypt the evidence using the proof key. We termed the decryption process of each party as ``plea of innocence'' since each party uses this process to prove that he is innocent.  Suppose that a receiver $R$ identifies a message $m$ being a false message and wants to find the transmitter $T$. Once the identity disclosure request has been approved by the majority, the identity disclosure process will be launched.   Recall that we assume that parties have the abilities to identify false information. A request can be approved only when most parties has identified the information is a false one. As shown in Fig.~\ref{fig:trace}, we use the former example to elaborate on the identity disclose process below:

\begin{algorithm} 
\KwData{$m_{fake}$, $EV_{i}$}
\KwOut {$R_{x}$} 

$j=i-1$;\\
$V_{j}= m_{fake}$;\\

\For {j \textgreater  0}   
{
\textbf{Relaying node $R_{j-1}$} release his proof key $PK_{R_{j-1}}$: \\

\If{$PK_{R_{j-1}} == \emptyset$}{

$R_{x}=R_{j-1}$\;
 RETURN  $R_{x}$\;

}\Else{

$  \textbf{DEC}(PK_{j-1},EV_{j})$ \\
$= {SIGN}(SK_{R_{j-2}}, ~\textbf{SIGN}(SK_{R_{j-1}} EV_{j-1} || V_{j})$;\\
$=S_{j}$
 
\If{$\textbf{VERIFY}(PubK_{j-1}, S_{j}, (V_{j} || EV_{j-1})))$}
{
\If{$\textbf{VERIFY}(PubK_{j-2}, S_{j}, (V_{j} || EV_{j-1})))$}
{

j=j-1\;

}\Else{
    $R_{x}=R_{j-2}$\;
 RETURN  $R_{x}$\;

}

}\Else{
    $R_{x}=R_{j-1}$\;
 RETURN  $R_{x}$\;

}

 }
 
}
 
$R_{x}=R_{j}$\;
 RETURN  $R_{x}$\;

\caption{ Identity disclosure protocol} \label{alg:2}
\end{algorithm}

\begin{enumerate}
\item The receiver $R$ requires a party $C$ to perform the plea of innocence since $R$ receives the false message from $C$. To prove so, $R$ locates evidence $EV_{4}$ on Blockchain, and makes the location of $EV_{4}$ and the $PK_{C-R}$ publicly accessible. In this way, all parties in the Blockchain P2P network can perform the decryption process, as show in equation (\ref{eq:dec}).
\begin{equation}
\begin{aligned}
 \textbf{DEC}(~PK_{C-R},EV_{4})  
 =~\textbf{SIGN}(SK_{R}, \\ ~\textbf{SIGN}(SK_{C}, m || EV_{3}  ))
 =S_{4}
\end{aligned}
\label{eq:dec}
\end{equation}
After the decryption process, all parties know that the previous hop is $C$, which was confirmed by $C$ and $R$, since their signature are present. As shown in equation \ref{eq:verify}, all parties can perform their signature verification without any changes\footnote{The verification process is similar to that of $R$, and we will not repeat it. }. 
\begin{equation}
\begin{aligned}
   ~\textbf{VERIFY}(PubK_{C}, S_{4}, (m || EV_{3})))
\end{aligned}
\label{eq:verify}
\end{equation}
 
\item In this case, $C$ is required to perform the plea of innocence. Different from the first step, the evidence $EV_{3}$ is contained in the plain-text of $EV_{4}$. So $C$ is only required to make the $PK_{B-C}$ publicly accessible. In such a way, parties know that previous hop is $B$, which was confirmed by $B$ and $C$, since their signature are present. Meanwhile, new evidence $EV_{2}$ shows up. 
\item Thereafter,  $B$ is required to perform the plea of innocence. Similar to the previous process, $B$ finally reveals a piece of new evidence  $EV_{1}$. 
\item Afterwards,  $A$ is required to perform the plea of innocence. The process is also similar to the  previous one. Finally, a a piece of new evidence $EV_{0}$ shows up. 
\item Finally,  $T$ is required to perform the plea of innocence.  $A$ is required to make a proof key publicly accessible. However,  $T$ does not have a proof key for $EV_{0}$, since $T$ is the message transmitter. Therefore,  $T$ is accused of false message spreading. In this case, $T$ can also release the three keys, which is considered as a ``confession''.  
 
\end{enumerate}

\subsection{Onionchain Based Vehicular Ad Hoc Networks}

We then introduce the work-flow of our Onionchain-based Vehicular Ad Hoc Networks (OVANETs), avoiding vehicles to spread false information intentionally. The motivation of this section is to demonstrate the generality of our Onionchain.

From a high level, OVANETs is built on a Blockchain-based reputation system \cite{li2012reputation,haddadou2014job}. In such a reputation system, vehicles share useful information, such as traffic information or information about road conditions, to gain their reputations. They are self-motivated by a suite of incentive mechanism. For example, vehicles with higher reputation may have more priority to access resources, or they can use reputations to earn their gas, while a vehicle with a lower reputation may fail to access some services. Therefore, each vehicle values its reputation and tries to have a higher reputation by behaving honestly and offering good service for other vehicles. Basically, a Blockchain-based database provides endorsements for our system. The endorsements are tamper-proof due to the advanced features of Blockchain. 

In such a scenario, privacy-preserving is an indispensable requirement \cite{lu2011pseudonym}. Vehicles may require to remain anonymous for various reasons. For example, vehicles may be easy to be convinced by a message from a vehicle with higher reputation, while they may fail to believe information from a vehicle with a low reputation. In this case, the vehicles with a low  reputation, like a vehicle which has just joined the system, may never have a chance to earn its reputation. Therefore,  Onionchain offers opportunities for vehicles with lower reputations due to our privacy-preserving. On the other hand, Onionchain can also work inversely and identifies these vehicles spreading false information. Once identified, punishments are enforced for these dishonest vehicles. One punishment can be decreasing their reputations. Finally, considering the limited storage resource of vehicles, the designer can choose RSUs to deploy the Onionchain, while vehicles can communicate with RSUs via VANET. 

It can be observed that our Onionchain can be extended to other similar scenarios with a little hindrance. For example, in the context of crowd-sourcing systems, employers/employees can also set up such an Onionchain based system to achieve privacy and traceability simultaneously. They may also use the idea of a reputation system to evaluate an employer/employee, and penalize the malicious parties when un-honest behaviors are detected. Moreover, by using the smart contracts introduced in section \ref{subsec:sc}, the entire process may be executed automatically without the human's involvement, reducing the burdens of management.

\section{Security Analysis}
\label{sec:secanalysis}

In this section, we will give the security analysis of our Onionchain. Specifically, an attacker may intentionally create a craft-packets and try to deploy attacks on our Onionchain in various ways. 
Notably, we examine the transmitter, relaying nodes, receiver throughout the life-cycle of a transaction and conduct five attack vectors in the regards of security analysis, as shown in Table ~\ref{tab:analysis}.  It can be observed that our Onionchain can hinder all the attack vectors without any change. 

\begin{table}  
\centering 
\caption{Attack Surface(\checkmark refers to our Onionchain can defend the attack without any changes)}
\label{tab:analysis}
\begin{tabular}{|P{1.5cm}|P{1.5cm}|P{1.5cm}|P{1.2cm}|}
 \hline
{\bf Attack Name} &  {\bf Initiator  }  &  {\bf  Number of Attackers } &  {\bf Onionchain }  \\
 \hline
Malicious-Transmitter  &  Transmitter  &   Single & \checkmark \\
 
Malicious-Messenger  &  Relaying nodes &  Single  & \checkmark \\
 
Replay  &  All participants  &    Single   & \checkmark  \\
 
 Calumniating  &  Receiver &    Single  & \checkmark \\
 
 Collusion  &  All participants &    Multiple  & \checkmark \\
 \hline
\end{tabular}

\end{table}

\textbf{Malicious-Transmitter Attack}:  In this type of attack, a malicious transmitter may create a false message intentionally, and uses another message, which is considered benign, to generate shreds of evidence. We assume that the false message is $m_{fake}$, and the benign message is $m$.  His motivation is to evade responsibility when the false message is detected. However, it is not possible for attackers to achieve so.  In this case, the party $A$ will not allow the transmitter to do so since the evidence is not the one $A$ received from the transmitter.  Even if $A$ is compromised by $T$, this type of attack still fails, since $A$ may require to publish all the keys eventually, and  $m_{fake} \neq m $. 
  
\textbf{Malicious-Messenger Attack}:  In this case, the malicious party is one of the relaying nodes, e.g. $B$ in our case.  
$B$ creates a false message $m_{fake}$ intentionally, and instead of using the original evidence, which is $ EV_{2}$ in our case, $B$ crafts new evidence $ EV^{\prime}_{2}$ based on a fake message $m_{fake}$.
He also crafts a fake $ V^{\prime}_{1}$ based on the fake message $m_{fake}$.
To notice, $B$ does not know who will be the receiver, so he chooses a receiver $R^{\prime}$ randomly. His goal is to conceive others to believe the false message is from the transmitter.
This type of attack will fail quickly. Since when the identity disclose protocol occurs, the attacker can not link the fake evidence $ EV^{\prime}_{2}$ to its previous evidence $ EV_{1}$. 
However, if he only crafts a fake $ V^{\prime}_{1}$ based on the fake message $m_{fake}$, and uses the original $EV_{2}$, the attack will still fail. The reason is similar to the first case, and we will not repeat. 

\textbf{Replay Attack}: Replay attack occurs when malicious relaying nodes resend a previous message and use the same evidence that used before. However, our timestamp can provide freshness for each message. The message receiver will discard the messages when they are stale.

\textbf{Calumniating Attack}: In this type of attack, an malicious receiver may create a false message/evidence intentionally, and tries to conceive others to believe the false message is from the transmitter. We assume that the false message is $m_{fake}$, and the original message is $m$. Therefore, in this case, $EV_{4}$ can be represented as follows: 
\begin{equation}
\begin{aligned}
 EV_{4} ~= \textbf{ENC}(~PK_{C-R},~\textbf{SIGN}(Sk_{R}, \\ ~\textbf{SIGN}(Sk_{C}, m_{fake} || EV_{3} ))) 
\end{aligned}
\end{equation}
However, the attacker can not modify or replace the evidence $EV_{3}$. This is because $ EV_{4}$ is generated by $C$ and $R$ corporately, and $C$ will not put his signature on it, if he detects $ EV_{3} \neq EV_{4}$. Therefore, as described in the identity  disclosure protocol, $EV_{3}$ can be traced back to $EV_{0}$. At this moment, $T$ will make the three keys, i.e $K_{T-A}$, $K_{A-B}$, $K_{B-C}$, publicly accessible, so that all parities can recover the original message $m$. It can be observed that $m_{fake} \neq m$, which turns out that $R$ tires to perform an Calumniating Attack.

\textbf{Collusion Attack}: In this case, two parties work jointly, and tries to craft a fake message. However, this type of attack is subject to the analysis in the previous examples. That is, the goal of an attacker is to create fake evidence or a fake message, and these fake ones fail to equal the previous evidence when parties perform the identity disclose. Therefore, our Onionchain can defend this type of attacks with no changes.  Moreover, in our paper, we demonstrate the case where there are only three relaying nodes are involved. It can be much more complicated for the attacker to deploy a Collusion Attack when more relaying nodes are presented, which hinders the Collusion Attack effectively. 
 
\section{Evaluation}
\label{sec:eva}
In this section, we evaluate our presented Onionchain. We first introduce the simulation environment of our prototype. We then introduce a few metrics to validate the feasibility and effectiveness of Onionchain. 

\subsection{Simulation Environment}
 
We simulate Onionchain as a Java Program (version 1.8). The Java Program communicate a local Ethereum Environment via web3j
Library and the smart contracts are programmed by Solidity language. Basically, the Oniochain client generates a transaction locally, and submits the transaction through the Web3j Interface, since Solidity language has limited support of cryptographic functionality. Also, we exclude the propagation delay, since it is negligible when compared with cryptographic operations, and it would be tricky to run our phototype on multiple machines.
In terms of cryptographic operations, we use AES-128 as our symmetric encryption algorithm and ECDSA as our signature generation/verification algorithm. We also assume that one party will perform our protocol instantly once he receives a specific packet. That is, we do not take human intervention as a consideration. This is reasonable since it is not possible to measure time cost when human intervention is involved.   Besides, The testing experiments are set up on a desktop computer with 2.20GHz Intel(R) Core(TM) Processor and 32GB memory.

\subsection{Numbers of Relay Nodes}

\begin{figure} 
\centering
\includegraphics[width=0.45\textwidth]{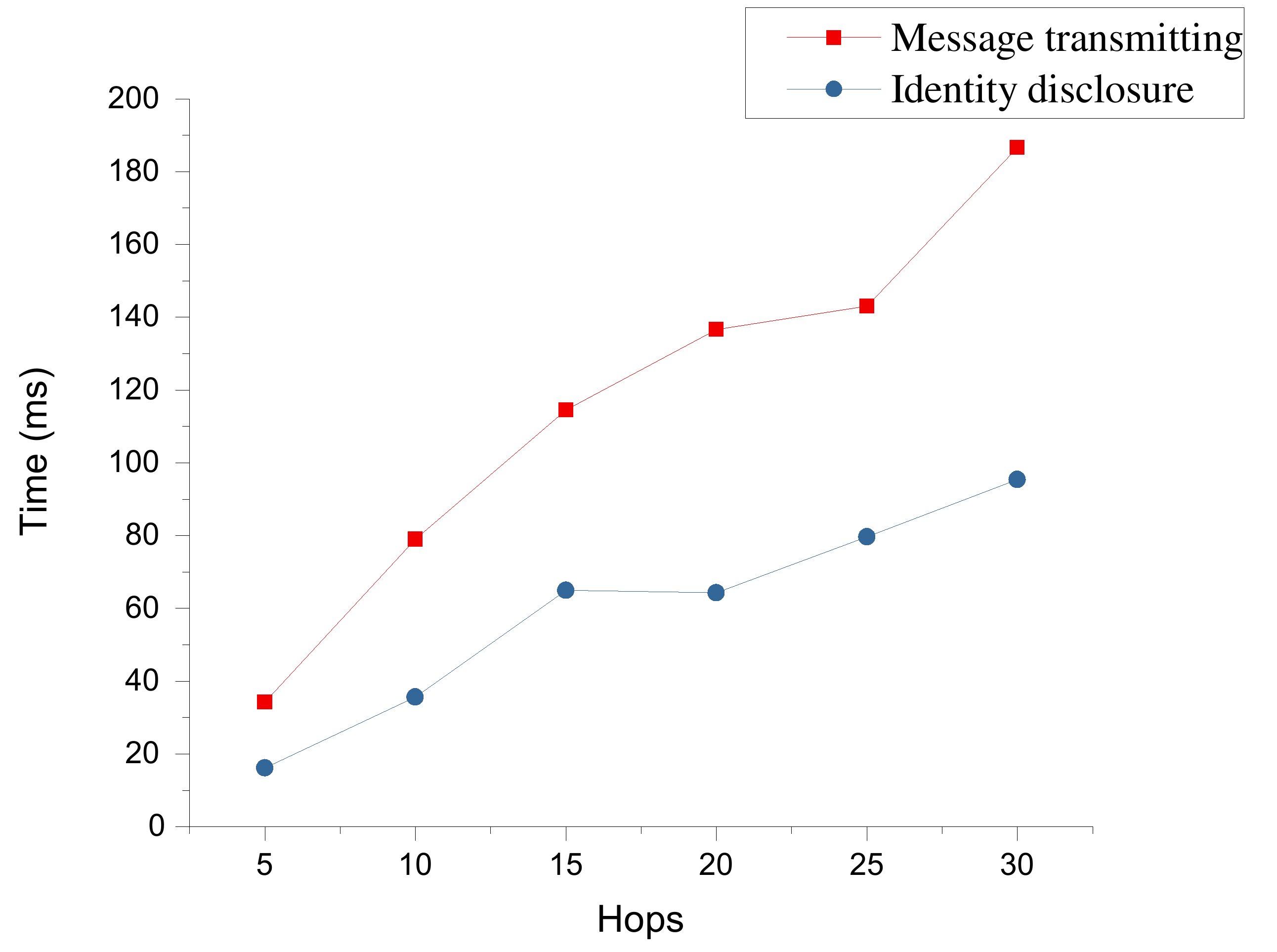}
\caption{\label{fig:time1} Numbers of relay nodes V.S. Time cost}
\end{figure}

Recall that the security of Onionchain partly depends on the numbers of relaying nodes. A transmitter may choose to use multiple relaying nodes to achieve better privacy. Fig~ \ref{fig:time1} shows the relationship between numbers of relaying nodes and the time cost. We choose different numbers of relay nodes and run our message transmitting protocol and identity disclosure protocol, respectively. Here, the message used to transmit via the network is 128 bit. It can be observed that all the two protocols cost less than 200,000 $u$s (i.e. 200 $m$s ), even if there are 30 relaying nodes are involved. Also, performing message transmitting protocol costs more time when compared with the other. This is because that message transmitting protocol uses more cryptographic operations. For example, message transmitting protocol involves signature generation while the identity disclosure is not. During our experiments, we also observe that when we execute one function twice during a short period of time, the second execution will be much faster than the first one. This may be because (i) if two calls are close to each other, the important information, such as the address of a function, might still be in the cache, meaning that the CPU does not need to fetch it again from memory. (ii) The hardware for executing the function is still active and ready after the first execution. This discovery will make more sense when we process a large number of messages in a real environment.

\begin{figure} 
\centering
\includegraphics[width=0.45\textwidth]{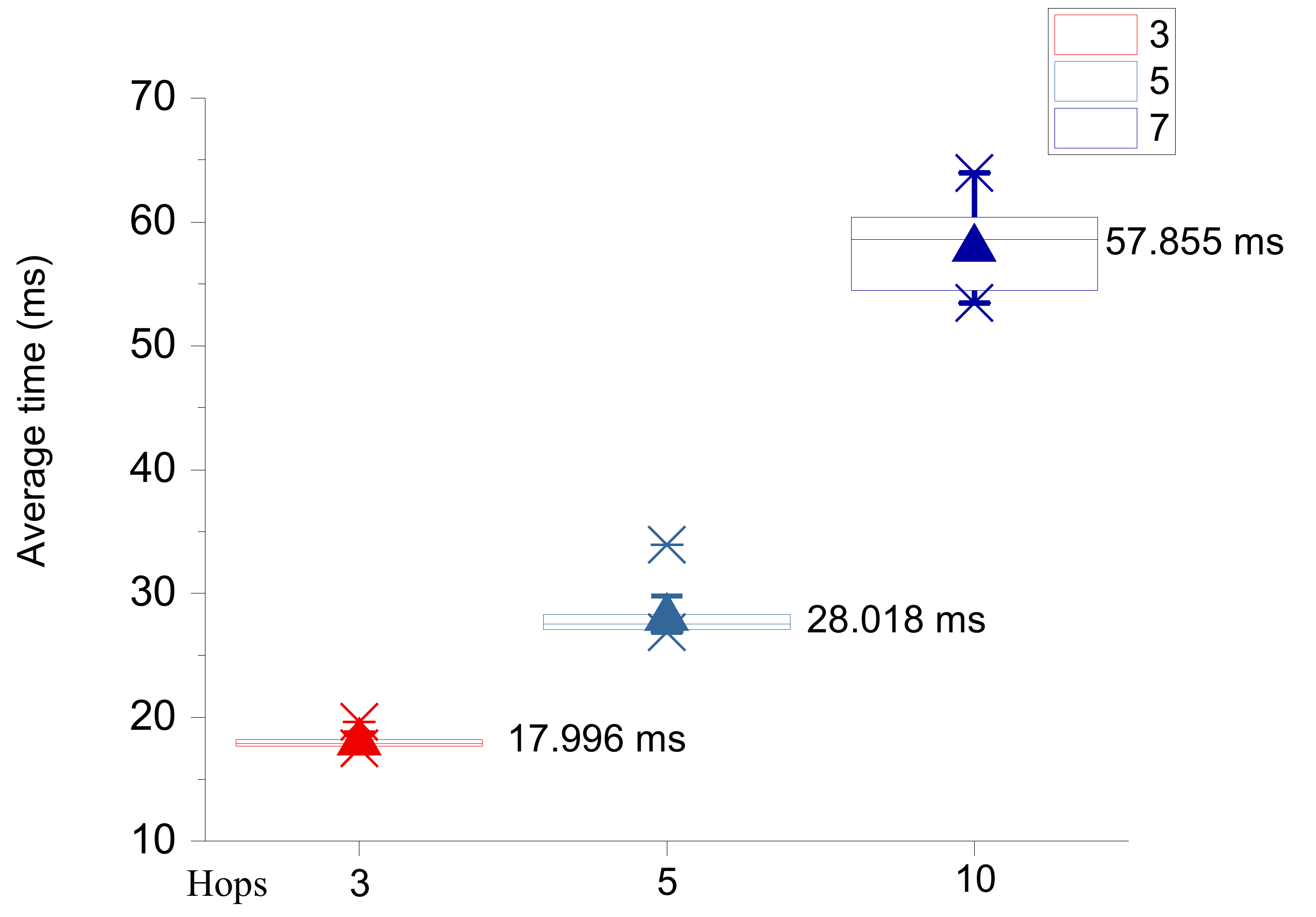}
\caption{\label{fig:time2} Average time cost of message transmitting}
\end{figure}

\begin{figure} 
\centering
\includegraphics[width=0.5\textwidth]{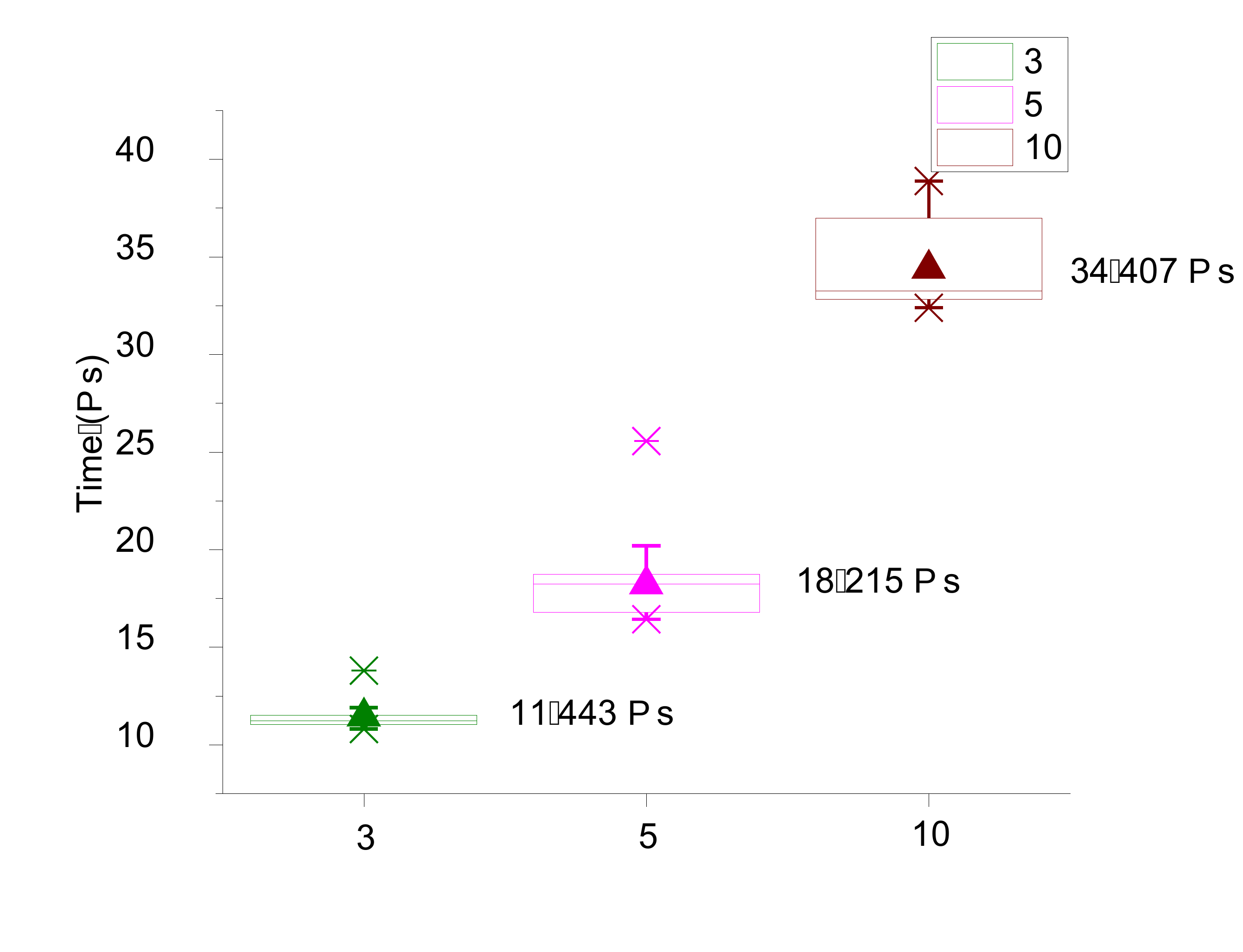}
\caption{\label{fig:time3} Average time cost of identity disclosure}
\end{figure}

\subsection{Average Time Cost}

We conduct experiments to evaluate Onionchain performance in regard of average time cost. To this end, we run the message transmitting protocol and identity disclosure protocol for 30 times and measure the time cost on average. Each time we use 3, 5, 10 relaying nodes respectively, to explore the performance in different privacy protection level.  Here, the message used to transmit via the network is 128 bit. It can be observed in Fig.~\ref{fig:time2} that the mean value of time cost of message transmitting is only 17996 $us$ (i.e. 17.996 $ms$), when there are 3 relaying nodes present. Even there are 10  relaying nodes involved, the mean value of time cost is still below 60 $us$, which is more than acceptable. As shown in Fig.~\ref{fig:time3}, the measurement from the identity disclosure protocol shows a better performance. It only costs around 10 $ms$ to perform, when there are 3 relaying nodes. When there are 10 nodes present, it costs around 35 $ms$ on average.

\subsection{Throughput}

\begin{figure} 
\centering
\includegraphics[width=0.45\textwidth]{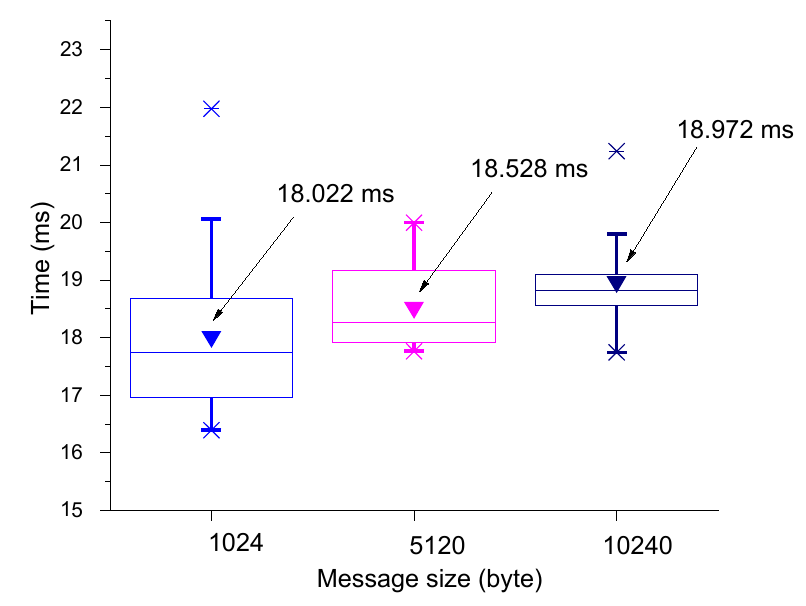}
\caption{\label{fig:time4}  Time cost for different size of packets in message transmitting}
\end{figure}

\begin{figure} 
\centering
\includegraphics[width=0.45\textwidth]{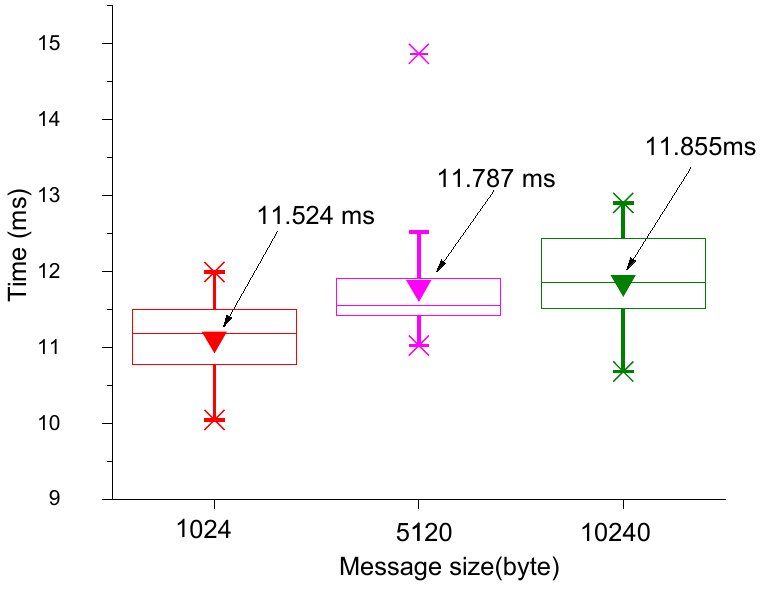}
\caption{\label{fig:time5} Time cost for different size of packets in identity disclosure}
\end{figure}

We now evaluate the time cost for different size of packets. Recall that in our previous experiment, the message used to transmit via the network is 128 bit. We choose this because a large packet may bring burden to the storage of Blockchain. Data expansion is still an open question that has not been well addressed for Blockchain applications.   
Even so, 128 bit may not be enough in exceptional circumstances, especially when files are involved. Therefore, we will explore the potential of Onionchian in terms of throughput. Fig.~\ref{fig:time4} and Fig.~\ref{fig:time5} show the results. It can be observed that even though the packets have grown to 10 Mbit, the overheads are still fairly small. The time costs have only increased less than 1 $ms$, in the cases of message transmitting and identity disclosure. We infer that this is because of the optimization of the run-time environment. For example, the function for encryption may remain active and ready for a long while, as long as it has been loaded into the memory.

\section{Related Work}
\label{sec:relatedwork}
In this section, we will review the research efforts that are related to ours.
In addition to what we have reviewed in the introduction, We will review more research efforts on Blockchain in the term of privacy and traceability. Afterwards, we then review  Blockchain-based vehicles communication systems.

We first review the privacy-preserving solutions related to Blockchain. 
In this regards, a lot of efforts have been made, while they fail to address remain the traceability. 
For example, 
Kosba \textit{et al.} present HAWK \cite{kosba2016hawk}, which stores the encrypted transactions instead of plain-text ones to protect transactional privacy. In addition, instead of presenting new solutions for privacy, there are a bulk of works that use Block as building blocks to address the privacy-preserving issues existing in a specific scenario. Zyskind \textit{et al.} \cite{zyskind2015decentralizing} use the Blockchain to manage personal data. In terms of privacy, the traditional third party is replaced by decentralized Blockchain system in their system, so that users will not have any concerns about the leakage of their privacy. 
Dorri \textit{et al.} \cite{dorri2017blockchain} discuss how to employ a Blockchain-based system to offer access control while preserving privacy in the context of  Internet of Things.
Ouaddah \textit{et al.} \cite{ouaddah2017towards} address a similar problem in IoT context using Blockchain. 
An extension has also been proposed by Dorri\cite{dorri2017distributed}, which discusses a few open questions for Internet-connected automotive in the regards of security and privacy. Yue \textit{et al.}\cite{yue2016healthcare} propose Healthcare Data Gateway, which addresses the privacy issues in the intelligence of healthcare systems. They use the Blockchain to manage and share data securely without violating privacy. A similar issue has also been addressed in Esposito's work \cite{esposito2018blockchain}.
Aitzhan \textit{et al.} \cite{aitzhan2016security} provide a Blockchain-based solution in the area of smart grid. Multi-signatures and anonymous encrypted messaging streams are involved in addressing transaction security in decentralized smart grid systems. A proof-of-concept is also implemented to validate their solution.  Liang \cite{liang2017provchain} propose a decentralized provenance architecture using blockchain technology, providing tamper-proof records, and achieving privacy of the provenance data in the cloud environment. 
 However, as discussed,  these Blockchain-based solutions are not omnipotent, since they may subject to powerful transaction analysis attacks \cite{reid2013analysis}.

 We now review the Blockchain-based solutions in the term of traceability. Tian \cite{tian2016agri} propose a Blockchain-based system to offer traceability for food supply chain, guarantee the food safety effectively.
 Another paper from him \cite{tian2017supply} demonstrate how the same idea can work in the food supply chain with HACCP (Hazard Analysis and Critical Control Points). Similarly, Lu \textit{et al.}\cite{lu2017adaptable}, Kamath \cite{kamath2018food} and Galvez \textit{et al.} \cite{galvez2018future} discuss how to trace the origin of products/food using  Blockchain. In addition to the food supply chain,  other applications are proposed based on the traceability of Blockchain. For example, Di \cite{di2018blockchain} \textit{et al.} use the Traceability of Blockchain to observe inter-organizational business processes.  The work that is most closely related to ours is the CreditCoin \cite{li2018creditcoin} by Li \textit{et al.} In their work, their goal is to provide a privacy-preserving Blockchain-Based incentive mechanism. They also discuss the traceability in their work. Our work is different from theirs. The reasons are (i) Their main concern is the incentive mechanism rather than traceability; (ii) They involve the Trace manager to trace malicious nodes. A trace manager is a group of decentralized parties other than regular users. In our paper, we do not make such an assumption. That is, every party can be the one who traces the transactions back. Therefore, our design is more realistic and decentralized; (iii) The generality of their design is not well discussed, while our work can adapt to many applications with little hindrance.

 We now review the vehicle communication systems that use Blockchain as their core engine. 
 Sean Rowan \textit{et al.}\cite{rowan2017securing} present a light inter-vehicle session key
establishment protocol using Blockchain technology, which establishes a trustworthy relationship between vehicles and manufacturers.
 Madhusudan Singh \textit{et al.}\cite{singh2017intelligent} propose an Intelligent Vehicle-Trust Point(IV-TP) mechanism using Blockchain technology, which provides security and reliability for vehicles behavior. Blockchain provides verifiable technical support for their solution. Similarly,  Sharma \textit{et al.} \cite{sharma2017block} build a transport
system based on Blockchain, allowing vehicles to achieve secure  
 Resources access. Jamin \textit{et al.} \cite{leiding2016self} deploy Blockchain technology in the context of a vehicular ad-hoc network (VANET), providing decentralized VANET services, including vehicle insurance, updates on traffic jams and weather forecasts. It can be observed that most of the efforts introduce Blockchain to solve a specific issue in the current vehicles communication systems.  Different from their works,  our onionchain uses the vehicles communication systems to demonstrate our generality.

\section{Conclusion}
\label{sec:con}
The fact that Blockchain technology fails to provide traceability and privacy-preserving simultaneously hinders Blockchain from developing and expanding its applications. To counter the issue, we propose Onionchain, offering applications a balanced option. Specifically, Onionchain integrates the mechanism of Onion routing into our protocol, enabling parties to perform transactions in an anonymous way. Besides, Onionchain records intermediate variables generated during the transactions, termed evidence, so that decentralized parties can trace a malicious party back when dishonest behaviors occur.   
An Onionchain-based Vehicular Ad Hoc Networks is demonstrated to provide community guidelines to follow, showing the generality of our design.  Extensive experiments are also performed to validate our Onionchain. 

We admit the current design also has limitations. For example, we do not take the scalability of blockchain into consideration\cite{karame2016security}. That is, our design also suffers from the transaction data explosion. To such a concern, we argue that (i) scalability of Blockchain is an open problem, which is out of our focus. 
(ii) The solutions to this issue have been widely discussed in previous works, such as \cite{poon2016bitcoin,eyal2016bitcoin}. Our approach does not retrofit the ecosystem of Blockchain. Therefore, these approaches can be integrated into our design with little efforts. (iii) For many Blockchain applications such as VANET, the Blockchain is deployed in RSUs or other Infrastructures, which have enough disk spaces. An infrastructure like RSU will have no concerns on the scalability of Blockchain. Finally, the next stage of our work falls into three aspects : (i) we may extend our work by exploring more scenarios. That is, more Onionchain based system will appear.  (ii) We may integrate solutions that address the scalability to refine our system.  (iii)  we may make our prototype public access, so that community can benefit from it in a timely manner. 
\section*{Acknowledgements}

Jian Weng was partially supported by National Key R\&D Plan of China (Grant Nos. 2017YFB0802203, 2018YFB1003701), National Natural Science Foundation of China (Grant Nos. 61825203, U1736203, 61732021), Guangdong Provincial Special Funds for Applied Technology Research and Development and Transformation of Important Scientific and Technological Achieve (Grant Nos. 2016B010124009 and 2017B010124002). Yue Zhang was partially supported by National Natural Science Foundation of China (Grant Nos. 61877029). Jiasi Weng was partially supported by National Natural Science Foundation of China (Grant Nos. 61802145, 61872153). Ming Li was partially supported by National Natural Science Foundation of China (Grant Nos. 11871248, U1636209).


\ifCLASSOPTIONcaptionsoff
  \newpage
\fi

\bibliographystyle{IEEEtran}
\bibliography{TDSC-Onionchain2019}
 \begin{IEEEbiography}[{\includegraphics[width=1in,height=1.25in,clip,keepaspectratio]{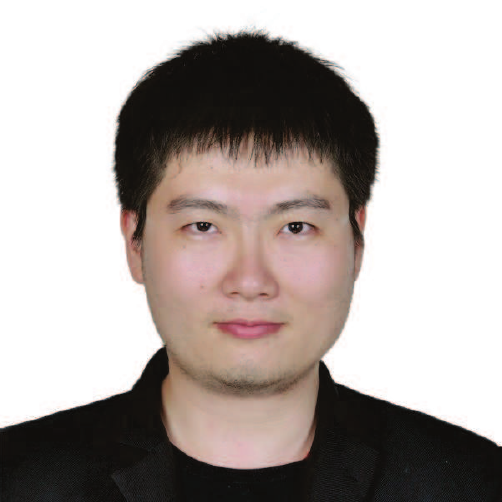}}]{Yue Zhang}
received his B.S. in  information security from  Xi'an University of Posts \& Telecommunications in 2013, and M.S. in information security from  Xi'an University of Posts \& Telecommunications in 2016. From 2016, he started his Ph. D. at Jinan University. His research interests include Bluetooth, system security and Android security.
He has published papers in international conferences and journals such as IEEE TDSC, IEEE TPDS, RAID etc.
\end{IEEEbiography}
 
\begin{IEEEbiography}[{\includegraphics[width=1.1in,height=1.25in,clip,keepaspectratio]{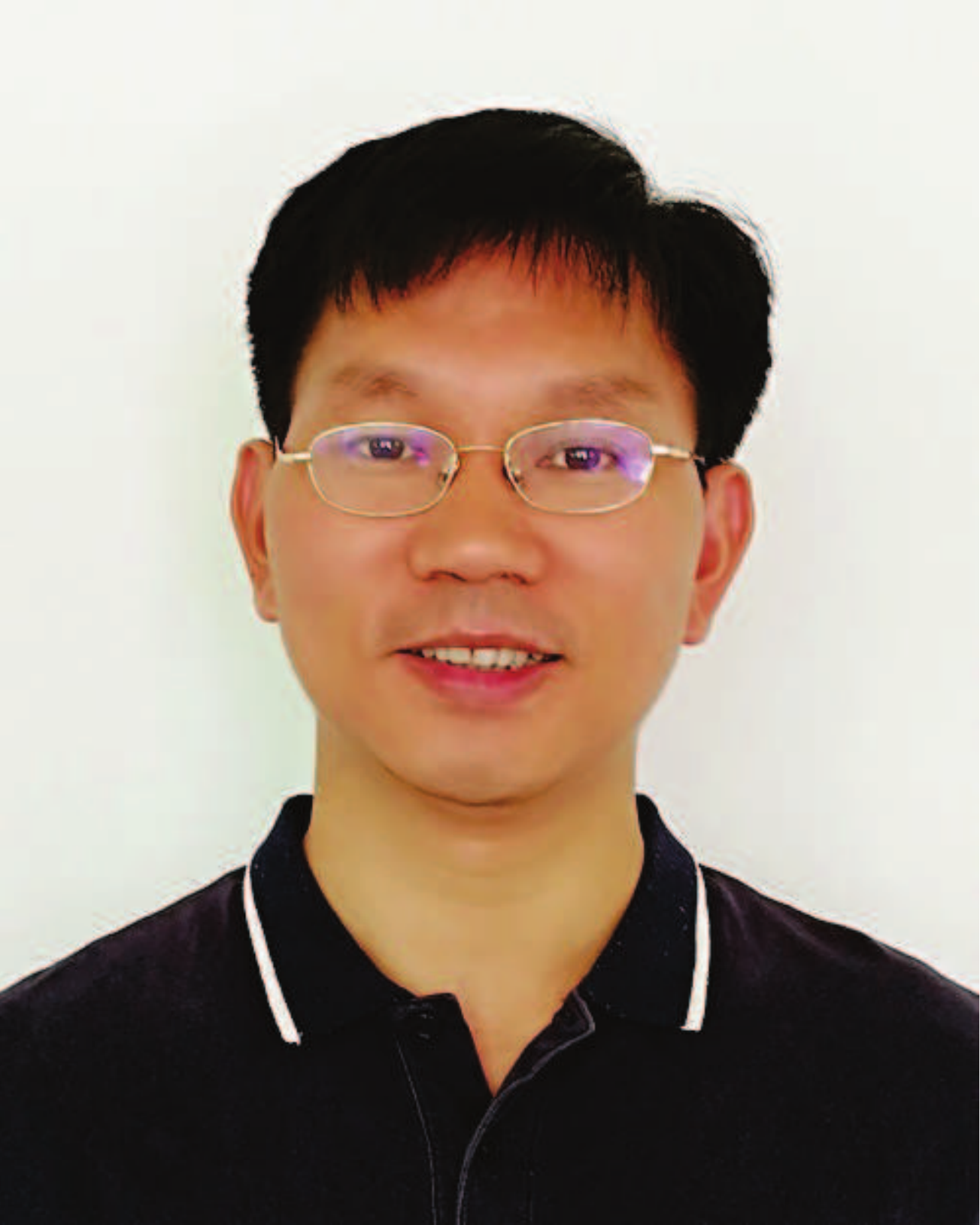}}]{Jian Weng}
is a professor and the Executive Dean with College of Information Science and Technology in Jinan University. He received B.S. degree and M.S. degree at South China University of Technology in 2001 and 2004 respectively, and Ph.D. degree at Shanghai Jiao Tong University in 2008. His research areas include public key cryptography, cloud security, blockchain, etc. He has published 80 papers in international conferences and journals such as CRYPTO, EUROCRYPT, ASIACRYPT, TCC, PKC, CT-RSA, IEEE TDSC, etc. He also serves as associate editor of IEEE Transactions on Vehicular Technology.
\end{IEEEbiography}
 
\begin{IEEEbiography}[{\includegraphics[width=1in,height=1.25in,clip,keepaspectratio]{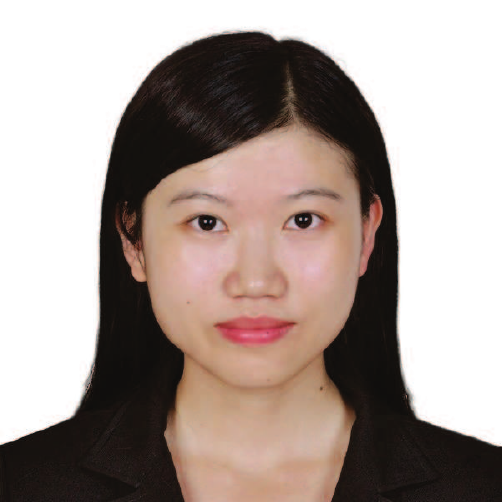}}]{Jiasi Weng}
 obtained the B.S degree in Software
engineering from South China Agriculture University in June 2016. She became a graduate student in Technology of Computer Application from Jinan University in September 2016. Her research
interests include cryptography and information
security, Blockchain and security in Software Defined Network..
\end{IEEEbiography}
 
\begin{IEEEbiography}[{\includegraphics[width=1in,height=1.25in,clip,keepaspectratio]{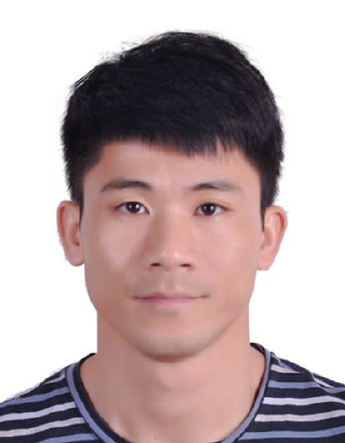}}]{Ming Li}
received his B.S. in electronic information engineering from University of South China in 2009, and M.S. in information processing from Northwestern Polytechnical University in 2012. From 2016, he started his Ph. D. at Jinan University. His research interests include crowdsourcing, blockchain and its privacy and security.
\end{IEEEbiography}
 
 \begin{IEEEbiography}[{\includegraphics[width=1in,height=1.25in,clip,keepaspectratio]{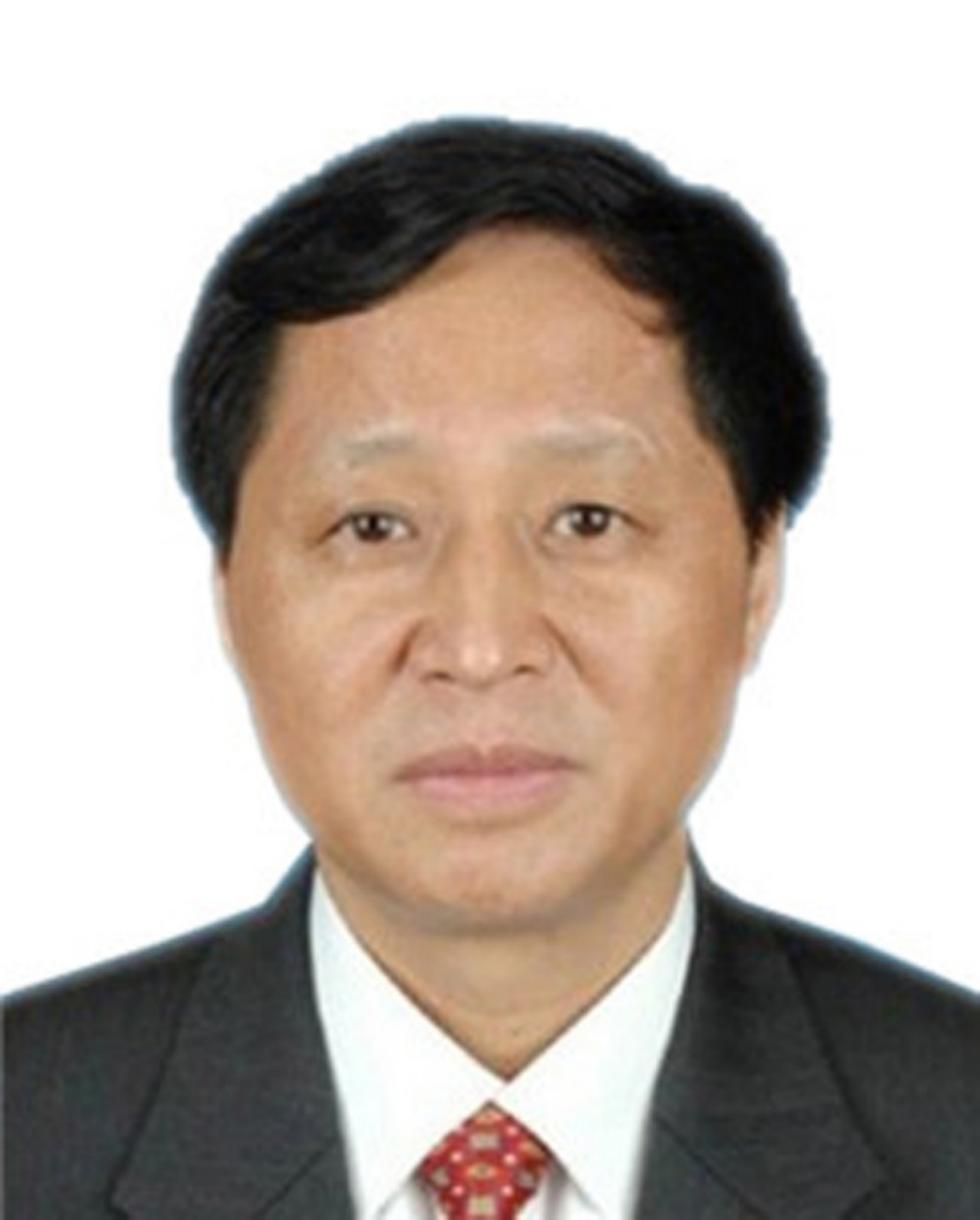}}]{Weiqi Luo}
  received his B.S. degree and M.S.
degree from Jinan University in 1982 and 1985
respectively, and Ph.D. degree from South China
University of Technology in 1999. Currently, he is
a professor with School of Information Science
and Technology in Jinan University. His research
interests include network security, big data, artificial intelligence, etc. He has published more than
100 high-quality papers in international journals
and conferences.
\end{IEEEbiography}

\end{document}